\documentstyle[epsf,aps]{revtex}
\renewcommand{\thefootnote}{\fnsymbol{footnote}}
\begin{document}
\newcommand{\be}{\begin{eqnarray}}
\newcommand{\dlq}{\lq\lq}
\newcommand{\ee}{\end{eqnarray}}
\renewcommand{\baselinestretch}{1.0}
\newcommand{\as}{\alpha_s}
\def\eq#1{{Eq.~(\ref{#1})}}
\parindent=20pt
\begin{flushright}
BNL-NT-99/7 \\
TAUP-2613-99
\end{flushright}
\vspace*{1cm} 
\setcounter{footnote}{1}
\begin{center}
{\Large \bf Large Scale Rapidity Correlations in Heavy Ion Collisions
} \\[1cm] Yuri V.\ Kovchegov$^1$, Eugene Levin$^{1,2}$ and Larry
McLerran$^1$ \\~~ \\ {\it $^1$Physics Department, Brookhaven National
Laboratory \\ Upton, NY 11973, USA \\~~ \\ $^2$ School of Physics and
Astronomy, Tel Aviv University \\ Tel Aviv, 69978, Israel
}\renewcommand{\thefootnote}{\fnsymbol{footnote}}\setcounter{footnote}{0}
\footnote{Permanent address} \\ ~~ \\ ~~ \\
\end{center}
\begin{abstract}
We discuss particle production mechanisms for heavy ion collisions. We
present an argument demonstrating how the fluctuations of the number
of produced particles in a series of classical emissions can account
for KNO scaling. We predict rapidity correlations in the particle
production in the event by event analysis of heavy ion collisions on
the rapidity scales of the order of $1/\alpha_s$.
\end{abstract}

\newcommand{\stackeven}[2]{{{}_{\displaystyle{#1}}\atop\displaystyle{#2}}}
\newcommand{\lsim}{\stackeven{<}{\sim}}
\newcommand{\gsim}{\stackeven{>}{\sim}}

\section{Introduction}

In ultrarelativistic heavy ion collisions in the upcoming experiments
at the Relativistic Heavy Ion Collider (RHIC) at Brookhaven and Large
Hadron Collider (LHC) at CERN, a very high multiplicity of particles
will be produced. Most of the models predict the multiplicities of
produced hadrons per unit of rapidity in a given event to be of the
order of \cite{RHIC}
\be\label{mult}
	{{d N} \over {d y} } \, \sim \, 10^3 , 
\ee
where the exact number depends on the particular model.  For these
high multiplicities the scale of statistical fluctuations in one unit
of rapidity is expected to be of the order of $\sigma \sim
\sqrt{dN/dy}$ so that
\be
	{{\sigma} \over {{d N}/{d y}}} \, \sim \, 10^{-1} - 10^{-2}.
\ee
Thus the multiplicities would not change much within each narrow bin
in rapidity, as long as the number of particles produced in that
rapidity bin is large. Collisions may therefore be studied on an event
by event basis with little ambiguity. The fluctuations of
multiplicities in each of the rapidity bins could be observed by
examining several different events. One might wonder whether the
numbers of particles produced in different rapidity bins are
completely independent of each other, or there are certain
correlations among them. In this paper we are going to address the
question of rapidity correlations in the framework of
McLerran--Venugopalan model of particle production by a series of
classical emissions \cite{mv,yv,KMW,JKMW,JKLW}.

\begin{figure}
\begin{center}
\epsfxsize=10cm
\leavevmode
\hbox{ \epsffile{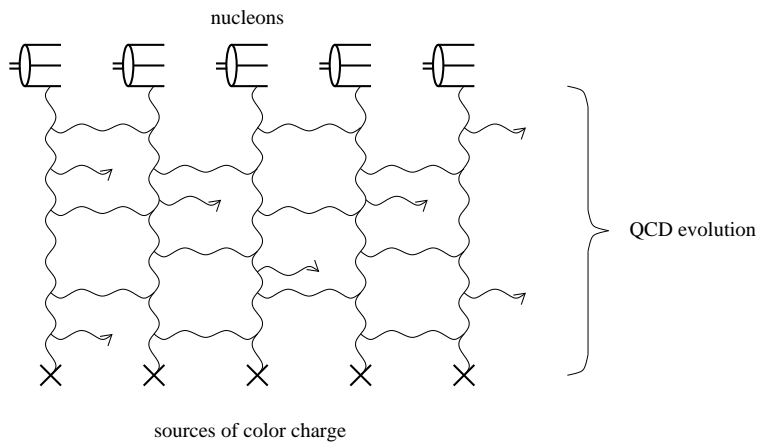}}
\end{center}
\caption{ An example of generation of color charge sources through the 
QCD evolution in a nucleus. The multiple reggeon evolution, which is
described in the text, after a number of iterations, that also include
emissions of gluons, produces a number of sea gluons in the rapidity
region far away from the nucleus, which later on act as sources of
color charge for classical gluon field. The sources are denoted by
crosses.}
\label{evol}
\end{figure}

In McLerran--Venugopalan model \cite{mv} the high multiplicity per
unit area gives rise to a dimensionfull parameter characterizing the
nuclear collision
\be\label{lam}
	\Lambda^2 = {1 \over {\pi R^2}} {{d N} \over {d y}}
\ee 
with $R$ the nuclear radius. It was argued that this scale represents
the typical transverse momentum of the particles produced in the early
stages of the nuclear collisions \cite{mv,mu}. Due to the high
multiplicities of Eq. (\ref{mult}) this typical transverse momentum
scale of produced partons given by Eq. (\ref{lam}) may become large
compared to the QCD scale
\be
  	\Lambda^2  \gg \Lambda^2_{QCD}.
\ee
The QCD coupling at the scale $\Lambda$ is weak, $\alpha_s (\Lambda)
\ll 1$. Together with the fact that the number of color charge sources
in the nuclei is large this allows us to assume that the gluons
produced in a nuclear collision could be described by the classical
field of the colliding nuclei \cite{mv,yv}. Here by classical field we
mean that the gluon field is a solution of Yang--Mills equations of
motion with the nuclei providing the source term for the equations
\cite{mv,yv,KMW}. A renormalization group (RG) approach has been developed 
recently to account for a series of classical emissions
\cite{JKMW,JKLW}. The procedure invented in \cite{JKMW,JKLW} is the
following: at each step of the evolution in rapidity each nucleus
through classical emissions produces several partons with soft
longitudinal momenta. At the next step of the RG the partons produced
in the previous step are included in the source, off which the next
generation of even softer partons is emitted. That way the produced
soft partons modify the density of color sources, for which one writes
a renormalization group equation \cite{JKMW,JKLW}. The exact form of
this equation is not important for our discussion here. We just note
that at the lowest order it reproduces the well known Balitsky, Fadin,
Kuraev, Lipatov (BFKL) equation \cite{BFKL} and the full equation
should be able to resum multiple reggeon exchanges in the structure
functions. The crucial assumption which we have to keep in mind is
that the series of classical emissions in both nuclei provide us with
a set of color charge sources, which give rise to the color field of
the produced gluons. This is illustrated in Fig. \ref{evol}. The QCD
evolution, which consists of real terms (classical emissions) and
virtual corrections provide us with the color sources, which are
denoted by crosses in Fig. \ref{evol}.

\begin{figure}
\begin{center}
\epsfxsize=6cm
\leavevmode
\hbox{ \epsffile{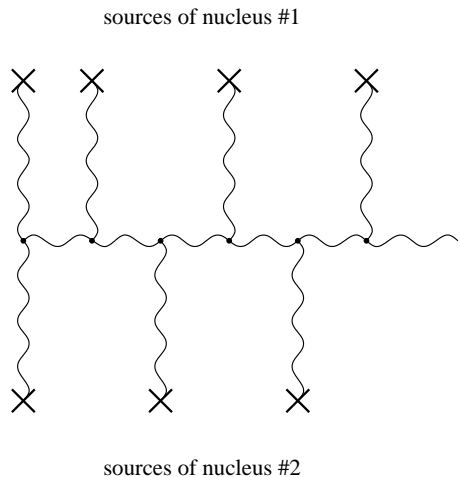}}
\end{center}
\caption{A classical gluon field produced in a nuclear collision as 
envisaged in the text. Through QCD evolution nucleons in nuclei give
rise to color charge sources (partons) denoted by crosses (see
Fig. \ref{evol}). The sources interact with each other to produce the
gluon field. }
\label{clf}
\end{figure}

To construct the classical field one has to solve Yang-Mills equations
treating the color charges generated through evolution of
Fig. \ref{evol} in both nuclei as contributions to the source term in
the equation. For the field to be purely classical the generated color
sources should be separated by the rapidity interval $\delta y \le
1/\as$. An example of the diagram contributing to the classical field
generated this way is given in Fig. \ref{clf}. The classical field of
colliding nuclei corresponds to gluon production in the approximation,
where one resums all powers of the parameter $\alpha_s^2 L$ \cite{yv},
where $L$ is the number of sources at a given impact parameter. If the
total rapidity interval between the colliding nuclei is not very
large, $Y \sim 1/\alpha_s$, so that the quantum evolution has not
become important yet, the valence quarks in the nucleons will be the
sources of color charge and the resummation parameter will become
$\alpha_s^2 A^{1/3}$ \cite{yv}.  One would have the nucleons instead
of the crosses in Fig. \ref{clf}. Finding the classical field even for
this somewhat simplified situation appears to be a complicated
task. The problem has been solved analytically only for the gluon
production in the case of a proton scattering on a nucleus in
\cite{mM}.  Several attempts have been made for the case of gluon
production in nucleus--nucleus collisions \cite{KMW}, giving only the
lowest order in $\alpha_s$ result. There have also been done extensive
numerical studies of the nucleus--nucleus collisions
\cite{raju}. In this paper we will not be interested in the detailed
structure of the field. What will be important to us is the fact that
the classical field is boost invariant, and, therefore, is rapidity
independent. Thus if one fixes the configuration of sources in
Fig. \ref{clf} the classical field in the rapidity interval $\Delta y
\sim 1/\alpha_s$ would give rise to a rapidity independent
distribution of the produced particles within this rapidity interval.

The classical gluon field has another interesting property. In the
regime when the number of color sources is large, making the
resummation parameter seizable, $\alpha_s^2 L \sim 1$, the classical
field of Fig. \ref{clf} becomes strong, $A_\mu \sim 1/g$
\cite{mv,yv,mu,mM}. Since gluons are massless bosons and the typical
phase space density is large, we expect that the gluon distribution
function at momentum scales less than $\Lambda$ will saturate with
occupation number
\cite{mv,yv,KMW,JKMW,JKLW,mu,mM,raju,Mueller1,Mueller2,Mueller3,MZ,glr}
\be\label{mul}
    {1 \over {\pi R^2}} {{d N} \over {d^2p_T d y}} \, \sim \,
    1/\alpha_s .
\ee
That way the field is very strong and non-perturbative, even though it
could be obtained in the weak coupling regime by usual perturbative
methods. The corresponding multiplicities are large [Eq. (\ref{mul})].

If the field is classical in each collision, the produced particles
arise from a fixed source on an event by event basis.  Each event is
characterized by a certain configuration of the color charges in the
source, off which the classical field of Fig. \ref{clf} is
emitted. The source results from the fluctuations in the color charge
density of those quarks and gluons at rapidities larger or smaller
than that at which we measure the distribution of produced
particles. We know that the spectrum of fluctuations is Poissonian for
a coherent state corresponding to a fixed source density which
produces the field. This is similar to the results of reggeon
calculus: after fixing the sources the classical fields might be
considered as reggeons independently producing particles (see
Fig. \ref{eve}). The spectrum therefore will be Poissonian
\cite{LRN,REGGEON,REG1,REG2,REG3,REG4,REG5,REG6,REG7,REG8,REG9}.  
Moreover, the classical effective action in a random background field
of \cite{JKMW,JKLW} describes both the dynamics of the gluon fields
and the fluctuations induced by changing the source strength
itself. Therefore using the model of classical emissions
\cite{mv,yv,JKMW,JKLW} we should be able to predict the spectrum of
rapidity fluctuations at particle formation time.

The nature of the fluctuations in the source density is complicated by
the fact that the source density itself satisfies a renormalization
group equation \cite{JKMW,JKLW}, and is correlated with what the
source would have been at other values of rapidity.  One of the
purposes of this paper is to disentangle this dependence and to
predict the general from of the multiplicity fluctuation spectrum.  In
Sect. II we shall show that as a consequence of the renormalization
group behavior, the fluctuation spectrum exhibits the Koba, Nielsen,
and Olesen (KNO) scaling \cite{kno}. The KNO scaling for
hadron--hadron collisions states that the probability of producing a
given multiplicity of particles $N$ at some given high energy $E$,
which we denote $d P / d N$, multiplied by the average multiplicity of
produced particles at this energy ${\overline N} (E)$ is described at
all energies by the same scaling function of $N /{\overline N} (E) $
\cite{kno}
\be\label{knohh}
\overline{N} (E) \,  {{d P} \over {d N} } = f(N / \overline{N} (E)),
\ee
characteristic of given hadron species. That way the energy dependence
of the high energy inclusive cross sections comes in only through the
average multiplicity of the produced particles ${\overline N} (E)$.

The nature of this scaling is slightly modified in the case of a
nucleus since in the nuclear collisions there is another parameter in
the problem --- atomic number of the nucleus $A$. Many nucleons in the
nucleus enhance multiple interactions, changing the shape of the KNO
distribution.  In this paper we will for simplicity consider only the
collisions of identical nuclei having the same atomic number $A$. The
relative width of the distribution depends on the baryon number of the
nucleus, and must scale roughly like $1/\sqrt{A^\delta}$, where
$\delta$ could be $2/3$, $1$ or $4/3$ depending on the saturation
model, as will be discussed below in Sect. II.  This scaling arises
from the fact that nucleons separated by transverse distances greater
than or of the order of a Fermi from one another must act
independently.  Thus for nuclear collisions we predict that the KNO
scaling function will depend on $A$
\be\label{knoaa}
\overline{N} (E) \,  {{d P} \over {d N} } = f(A, N / \overline{N} (E)).
\ee
We will derive this result in Sect. II.  It implies that for a nucleus
of fixed size $A$ one should observe KNO scaling, but the scaling
function will differ from a nucleus to a nucleus.  We will also argue
that the form of this distribution is
\be
	f \sim \exp{ \left[ - const \, A^\delta \, (\sqrt{N /
	\overline{N} (E)} - 1)^2 \right]}
\ee
[see Eq. (\ref{knof})]. This multiplicity distribution has a width
\be
	\delta N \, \sim \, \overline{N} (E)/\sqrt{A^\delta}.
\ee
Of course the width $\delta N$ is parametrically of the order of
$\sqrt{A^\delta}$, since, as will be shown in Sect. II,  ${\overline N}
(E) \sim A^\delta$, but the signal for the KNO form is its dependence
on energy through the total multiplicity.

Of course, due to the final state interaction of particles, the number
of particles in the final state is not expected to be much changed
from the initial conditions.  We expect therefore that fluctuations in
the total particle multiplicity as a function of rapidity on an event
by event basis at the particle formation time should be reflected in
the final state distribution of produced particles.  This should be
true up to at least some number such as $\alpha_s$ times
$\sqrt{dN/dy}$.  If we look for very rare fluctuations, then such
fluctuations in the final state interactions should not obscure a huge
initial state fluctuation.

In Sect. III we will derive another feature of the underlying
classical dynamics which manifests itself in the two particle
distribution function.  To compute the two particle distribution, for
each event we take the multiplicity in some bin at $y_1$ and multiply
by the multiplicity at another rapidity bin $y_2$.  If the classical
fields are the ones responsible for the distribution of the produced
particles, we shall show that at the leading order in $A$ the two
particle distribution function factorizes into
\be
	D (y_1,y_2) = {{d N} \over {d y_1}} \, {{d N} \over {d y_2}}.
\ee
The factorization is demonstrated in Fig. \ref{eve}.  Of course this
factorization would happen if the distributions were uncorrelated at
rapidities $y_1$ and $y_2$.  We shall argue that they are in fact very
tightly correlated in Sect. IV by introducing a correlation function
which would allow one to disentangle the classical effects from other
possible mechanisms of particle production.

The classical field responsible for this distribution is constant over
a large rapidity interval.  Therefore if we measure the multiplicity
fluctuation in a rapidity interval centered around $y_1$ which is
several sigma away from the average, the multiplicity in a neighboring
rapidity interval centered around $y_2$ should be roughly the same
multiple sigma fluctuation from the average.  This very remarkable
correlation is a measure of the classical coherence of the field, and
is a direct measure of the underlying strong like dynamics of the
gluon field.  We shall make this argument firm in Sect. IV by
introducing a correlation coefficient at a fixed impact parameter of
the nuclei ${\cal C} (b, y_1, y_2)$ in Eq. (\ref{ccoef}). The
coefficient ${\cal C} (b, y_1, y_2)$ is an experimentally measurable
observable which is equal to $1$ when the particle productions at the
rapidities $y_1$ and $y_2$ are correlated on the event-by-event basis,
and is less than one otherwise. We will argue that if the classical
picture of emissions is true this correlation coefficient should be
$1$ over large intervals in rapidity, $|y_1 - y_2| \sim
1/\alpha_s$. It should fall off for wider rapidity intervals. This
prediction could be checked experimentally at RHIC and LHC.

In Sect. IV we also discuss the differences and similarities of the
results of our classical approach and the conclusions for rapidity
correlations which could be driven out of the old reggeon theory.

We summarize our results in Sect. V.

\section{Nature of KNO Scaling}

In this section, we will study the fluctuation spectra of produced
particles.  Strictly speaking, we are studying the fluctuation
spectrum of produced gluons in early stages of the collisions.  The
number of gluons is of course modified by subsequent interactions with
other gluons and in their subsequent transmutation into pions.  It is
expected nevertheless that the number of produced pions is close to
the number of initially produced gluons. This is because the gluons
are thermalized largely by two body collisions in the early stage of
the collision, since the coupling is weak if the typical transverse
momentum of the gluons is large. This is expected if the multiplicity
per unit area $\Lambda^2$ of Eq. (\ref{lam}) is large compared to
$\Lambda^2_{QCD}$ as should be the case for large nuclei at
asymptotically high energy.  The two body collisions change the
transverse momentum distribution of gluons, but preserve the total
gluon number. In this paper we are interested in the multiplicities of
gluons integrated over all transverse momenta (or coordinates), which
do not change under thermalizing two gluon collisions.  At later times
after thermalization, we expect that entropy will be approximately
conserved.  At late stages, the entropy is converted into pion number
as the system cools.  We expect therefore that $dN_{gluon}/dy
\sim dN_{pion}/dy$.  In general there may be some weak dependence of this
proportionality upon multiplicity, but for our purposes this weak
dependence will not be important.

Therefore, although we compute the initial multiplicity fluctuations in
gluons, this should be reflected in the multiplicity fluctuations of
produced pions.

In the McLerran-Venugopalan model of the small x hadronic wave
function, one computes a classical field which arises from a source
density \cite{mv,JKMW}.  This source density is at rapidities much
larger than that of the classical field which we compute.  Eventually
the source density is itself integrated over to be included in the
source for the next series of classical emissions \cite{JKMW,JKLW}, as
was described in the introduction.

Let us first compute the fluctuation spectrum of produced particles
for a fixed source.  In this case, the wavefunction corresponding to
this classical field is a coherent state.  Coherent state wavefunction
generate Poisson distribution in multiplicities.  Therefore the
typical fluctuation scale in a unit of rapidity is of order
$\sqrt{dN/dy}$.  We will soon see that this scale is small compared to
that generated by the fluctuations in the source itself.

To understand the fluctuations induced by the source of the classical
fields, we construct a model which has most of the features of BFKL
evolution and fluctuations in the source density.  We use intuition
developed from understanding the renormalization group structure of the
McLerran-Venugopalan model.  We use the equation 
\be\label{stoc}
	{{d N} \over {d y}} = \kappa N(y) + \sqrt{\kappa' N(y)} \zeta
	(y).
\ee
In this equation, $N(y)$ is the total number of particles in the rapidity
range between $y_{target}$ and $y$,
\be
	N(y) = \int_{y_{target}}^y d y^\prime {{d N} \over {d y^\prime}}.
\ee
The first term on the right hand side of the evolution equation
(\ref{stoc}) is the toy model analog of the kernel of the BFKL
evolution equation (thus $\kappa = \alpha_P -1 = \frac{4 \alpha_s N_c
\ln 2}{\pi}$ is the intercept of the BFKL pomeron \cite{BFKL}). For a
series of classical emissions one could conclude that the number of
produced particles is proportional to the total number of partons off
which the new particles are emitted. This is reflected in the first
terms of Eq. (\ref{stoc}). We note that due to the classical emission
picture we can write Eq. (\ref{stoc}) for particle multiplicities and
not just for the cross sections (The BFKL equation was originally
written for the cross section.). The reason for that stems from the
fact that at the high energies considered here the total cross
sections become independent of energies, thus making the energy
dependence of inclusive cross sections and multiplicities identical.
If we ignore the second term on the right hand side of
Eq. (\ref{stoc}) the solution of the equation would be
\be\label{1pom}
 	{{d N} \over {d y}} = N_0 e^{ \kappa (y-y_{target}) },
\ee
which is our analog of the solution of the BFKL equation. Comparing
Eq. (\ref{1pom}) to the usual one BFKL pomeron exchange result
\cite{BFKL} we note here that $N_0 \sim \alpha_s^2$.  The $A$-dependence 
of $N_0$ is determined from the predictions for multiplicity inside of
the saturation region. All saturation models \cite{mv,glr,me} agree
that the multiplicity of the produced particles in the saturation
region is given by 
\be\label{sat}
{{d N} \over {d y}} \, \sim \, A^{2/3} \, Q_s^2 (y)
\ee
with $ Q_s (y)$ the saturation momentum, which is denoted by $\Lambda$
in \eq{lam}. The factor of $A^{2/3}$ in \eq{sat} results from the
integration over the impact parameters of the nuclei. In the
Glauber--Mueller--type saturation models \cite{Mueller4} the
saturation momentum scales as $Q_s^2 \sim A^{1/3}$ due to multiple
rescatterings within the nucleus. In the approaches based on the BFKL
equation \cite{mv,me,LT} $Q_s^2 \sim A^{2/3}$. In the Reggeon approach
\cite{LRN,REGGEON,REG1,REG2,REG3,REG4,REG5,REG6,REG7,REG8,REG9} $Q_s^2$ 
is independent of $A$. We can summarize all these results by writing
$N_0 \sim A^\delta$, with $\delta$ being dependent on the particular
model at hand. In Glauber--Mueller model $\delta = 1$, in BFKL-based
approaches $\delta = 4/3$ and in Reggeon calculus $\delta = 2/3$.

The second term on the right hand side of Eq. (\ref{stoc}) is
stochastic and in our model describes the fluctuations induced at each
stage of the evolution equation.  The probability distribution of the
fluctuation of the stochastic term $\zeta$ is Gaussian
\be\label{Z}
   Z = \int [d \zeta] \, e^{- {1 \over 2} \int d y \zeta^2 (y)  }.
\ee
Its origin is from the renormalization group \cite{JKMW}.  At each
step a classical field is induced, which when small fluctuations are
computed gets converted into a source for the next step in the
evolution.  Of course the classical field itself has Poissonian
fluctuations which near the center of the distribution should be
Gaussian.  Therefore, the induces source at the next step will have
fluctuations built in which are correlated with the fluctuations in
the previous step.  If we look over an interval of unit width in
rapidity, the weight of these fluctuations is of order $\sqrt{ dN/dy
}$.  Our stochastic source $\zeta$ is weighted by the function $Z$ of
Eq. (\ref{Z}) so that its fluctuations in one unit of rapidity are
also of order one.  Using the fact that $\kappa \, N \sim \, dN/dy$
which is true for the solution (\ref{1pom}) of Eq. (\ref{stoc})
without the stochastic term, and, as we shall see below, is
approximately valid for the full Eq. (\ref{stoc}), we can replace the
weight of the stochastic fluctuation term by $\sqrt{dN/dy} \rightarrow
\sqrt{\kappa' \, N}$ arriving at the evolution equation
(\ref{stoc}). We also note that $\kappa' \sim \alpha_s$.

The solution to Eq. (\ref{stoc}) for fixed $\zeta $ is
\be\label{eqsol}
	N(y) = \left[ \sqrt{\overline N(y)} + {1 \over 2}
	\sqrt{\kappa'} \int_0^y d y^\prime e^{\kappa (y-y^\prime)/2}
	\zeta (y^\prime ) \right]^2 .
\ee
In Eq. (\ref{eqsol})
\be
	\overline N (y) = e^{\kappa y} N_0
\ee
where we will in the future set $y_{target} = 0$.  Note that
$\overline N$ is the typical average multiplicity at fixed $y$ which
one would have in the limit when fluctuations are turned off.

We can now compute the multiplicity distribution function $dP/dN$.  To
this we must integrate over the fluctuations fields $\zeta$ with the
constraint that the total multiplicity is given by Eq. (\ref{eqsol}).
We have
\be\label{pn1}
	{{d P} \over {d N}} = \frac{1}{Z} \, \int [d\zeta] \exp\left[
	- {1 \over 2} \int d y^\prime \zeta^2(y^\prime) \right]
	~\delta \left( N(y) - [\sqrt{\overline N(y)} - {\sqrt{\kappa'}
	\over 2} \int_0^\infty d y^\prime e^{\kappa (y-y^\prime)/2}
	\zeta (y^\prime ) ]^2 \right)
\ee
In this integral, we replaced the upper limit of integration in the
expression for $N(y)$ by infinity, since we are typically interested in
large values of $y$.  This makes the analysis simpler in what follows.

To evaluate the path integral, one should decompose
\be
	\zeta(y) = \sum_{n=0}^\infty \, e^{-\kappa y/2} \, c_n \, L_n
	(\kappa y),
\ee 
where the $L_n$'s are Laguerre polynomials $L_n^0$.  Using the
orthogonality condition for Laguerre polynomials one can see that all
the integrals over $c_n$'s in Eq. (\ref{pn1}) become Gaussian with the
exception of the integral over $c_0$ in the numerator, which is fixed
by the delta function. The integrals over $c_n$'s for $n \ge 1$ can be
done in closed form and are canceled by the same integrals in $Z$ in
the denominator. The final answer is given by the $c_0$ integration.

In the end, we find that
\be\label{knoff}
	{\overline N} (y) \, {{d P} \over {d N}} = \sqrt{ {\kappa N_0
	{\overline N} (y) \over {2 \pi \kappa' N}} } \exp \left[ - 2
	\, \frac{\kappa N_0}{\kappa'} \left( 1 -
	\sqrt{\frac{N}{\overline N(y)}} \right)^2 \right].
\ee
The multiplicity of the produced particles $N_0$ is very large. That
allows us to expand \eq{knoff} around $N = {\overline N}$. We obtain
\be\label{knof}
	{\overline N} (y) \, {{d P} \over {d N}} = \sqrt{ {\kappa N_0
	\over {2 \pi \kappa'}} } \exp \left[ - {\kappa N_0 \over {2
	\kappa'}} \left( 1 - \frac{N}{\overline N(y)} \right)^2
	\right].
\ee
This result for the multiplicity distribution is simple to understand.
The evolution equation connects the produced particle to some initial
spectrum of fluctuations.  The relative importance of these
fluctuations decreases as the multiplicity increases.  Therefore the
fluctuations within an interval of width $1/\alpha_s$ in rapidity set
up the fluctuations at higher values.  This is the factor of $\kappa
N_0/\kappa'$ in the exponent of \eq{knof}. Remembering that $\kappa
\sim \kappa' \sim \as$ and  $N_0 \sim
\alpha_s^2 A^\delta$ we conclude that the relative width of the
multiplicity distribution scales like $1/\alpha_s
\sqrt{A^\delta}$. The fluctuations are all correlated all the way down
the chain and therefore should only depend upon the ratio $N/\overline
N(y)$.  This dependence is the essence of KNO scaling.  The only
important variable is the multiplicity divided by the average
multiplicity.  Of course there is also a dependence on $N_0$ because
there are more independent emitters for a nucleus than for a hadron.
This is reflected in the fact that $N_0 \sim A^\delta$.

Note that the width of this distribution is
\be\label{wd}
	\delta N^2 \sim \overline N^2(y) \kappa' \, / N_0 \kappa \, .
\ee
Parametrically $\delta N^2$ is linear in $A^\delta$, as $N_0\sim
A^\delta$ and ${\overline N} \sim N_0 \sim A^\delta$.

In general, the fluctuation spectrum for the first few emitters which
generate the fluctuations in the distribution may be different from
\eq{knof}. The exact shape of the KNO scaling function will probably
differ from the one given by Eqs. (\ref{knoff}) and (\ref{knof}).
Nevertheless, the physical picture we have generated still is true,
and therefore we expect that in general the distribution will be of
the form
\be
{\overline N} (y) \, {{d P} \over {d N}} = f(N_0, N/\overline N(y) )
\ee
and that the typical width of the distribution will scale with $A$ and
$\alpha_s$ in the same way as the width in our model given by
Eq. (\ref{wd}).  The physical picture we have of the small $x$ gluon
distribution functions therefore automatically has KNO scaling
[cf. \cite{Mueller3}].

In terms of the strong coupling constant the relative width $\delta
N^2 / {\overline N}^2 (y) \, \sim 1/\alpha_s^2$ (see
Eq. (\ref{wd})). It has been argued in the framework of
McLerran--Venugopalan model that in high energy nuclear scatterings
the high parton density sets the scale of the running coupling
constant \cite{mv,JKMW,JKLW}. If this assumption is true, than one
could conclude that at very high energies $N_0 \sim \as^2
(\Lambda_{target})$, since $N_0$ is the multiplicity of the particles
in the fragmentation region. $\Lambda_{target}$ is the transverse
momentum scale characterizing the target nucleus.  At the same time
$\kappa \sim \as (\Lambda)$, as it determines the evolution of the
multiplicity $N (y)$ at some large rapidity $y$. ($\Lambda$ is given
by \eq{lam}.) $\kappa'$ depends on $y$ through its dependence on
$\Lambda \sim N (y)$. However, if we allow $\kappa'$ to depend on $y$
then, since the integral over $y'$ in \eq{eqsol} is dominated by $y'
\approx 0$ one can see that the most important value of $\kappa'$
would be at the fragmentation region. Thus $\kappa' \sim \as
(\Lambda_{target})$. In the light of the above estimates and using
\eq{wd} we conclude that the relative width depends on the average
multiplicity
\be\label{knov}
\frac{\delta N^2}{{\overline N}^2 (y)} \, \sim  \frac{1}{\alpha_s (\Lambda) \,
\as (\Lambda_{target})} \sim \frac{1}{\as (N)}.
\ee
The high multiplicity of the produced particles would create a large
momentum scale $\Lambda$, which would make the coupling constant
$\alpha_s (\Lambda)$ small. Thus the relative width of the KNO
distribution given by Eq. (\ref{knov}) will get larger. Moreover,
since the change in width would depend on the average multiplicity of
the produced particles that would violate the KNO scaling. Thus our
prediction of KNO scaling will start to break down slowly at very high
energies. At the same time since the rate of change of the running
coupling constant $\as (\Lambda)$ slows down at large $\Lambda$ the
violation of KNO scaling will decrease as the energy gets higher.

\section{Classical Nature of the Distribution Function}

Another feature of the classical field particle production arises from
considering the distribution function $D (y_1,y_2)$.  We define this
to be the two particle distribution function measured in the following
way: In each event or class of events, measure the multiplicity of
particles at rapidities $y_1$ and $y_2$ in bins of width $d y_1$ and
$dy_2$.  Multiply these multiplicities together to generate
\be
	d^2 N(y_1,y_2) = D(y_1,y_2) \ d y_1 \, d y_2 .
\ee
Then average $D (y_1,y_2)$ over events or classes of events.

\begin{figure}
\begin{center}
\epsfxsize=8cm
\leavevmode
\hbox{ \epsffile{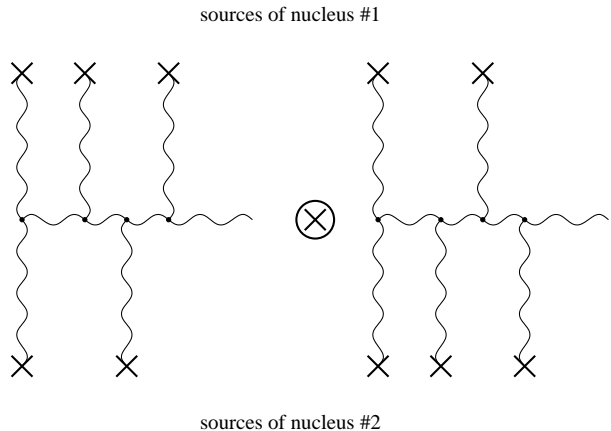}}
\end{center}
\caption{ Classical production of two gluons in nuclear collision as 
described in the text. The figure illustrates factorization of
\eq{cor}.}
\label{eve}
\end{figure}

A class of events might be generated by putting a cut on say the total
multiplicity of particles in the neighborhood of zero rapidity.  In the
class of events where this cut is satisfied, one could measure the
multiplicity in different regions of rapidity and average over the class
of events.

The distribution function $D$ may be generated as an expectation value of
the number operator
\be
	D(y_1,y_2) = \int {{d^2p_T} \over {(2\pi)^2}} {{d^2q_T} \over
	{(2\pi)^2}} \, \left< a^\dagger (y_1,p_T) \, a (y_1,p_T) \,
	a^\dagger (y_2,q_T) \, a (y_2,q_T) \right>.
\ee
In the classical field limit, this expression is of the form
\be\label{aaaa}
     \left< a^\dagger (y_1,p_T) \, a (y_1,p_T) \, a^\dagger (y_2,q_T)
     \, a (y_2,q_T) \right> \, \sim \, A^i(p) A^i(-p) A^i(q) A^i(-q)
\ee
where $A^i$ is the classical field produce by the color sources in the
colliding nuclei \cite{KMW,mM,raju}.  When one averages over the
sources which generate the fields, one finds that
\be\label{cor}
	D(y_1,y_2) = \left< {{d N} \over {d y_1}} \right> \, \left<
	{{d N} \over {d y_2}} \right>.
\ee
This is true up to corrections which are of order $A^{-2/3}$, that is
of the order of one over the area of the nuclei. This statement could
be understood from Fig. \ref{eve}. The classical fields producing the
gluon connect to several color sources. In the leading order in $A$
the fields generating gluons at $y_1$ and $y_2$ connect to different
sets of color sources, making Eq. (\ref{cor}) true.

Note that this tells us that the fields are essentially trivially
correlated in the longitudinal direction.  The connected piece of $D
(y_1,y_2)$ has vanished entirely, up to corrections which go like one
over the area of the nuclei.  This lack of correlation is not by
itself evidence of a classical field.  It could occur if there were
for example no correlations at all in the longitudinal space.  The
structure of the KNO distribution itself suggests that this is not the
case.  In the next section, we shall see that there is a correlation
function which dramatically illustrates the classical correlation.

\section{Classical Correlation}

If there are classical fields, then, as we have seen above, due to
boost invariance, these classical fields are independent of rapidity
over a wide range of rapidity \cite{KMW,mM}.  Therefore the
multiplicity should on the average be the same over the range of
rapidity where the classical field theory is valid.  This is typically
of order $1/\alpha_s$ in the classical field approach.  At the same
time $\alpha_s$ should be small when the density of produced particles
is large \cite{mv}, making this correlation length in rapidity large.

This effect can be measured in the following way: Measure the rapidity
density in some bin of width $dy_1$ around $y_1$ and $dy_2$ around
$y_2$.  Require that $dy_1$ and $dy_2$ are large enough so that the
statistical fluctuations in the rapidity in a given event are small
compared to the total multiplicity.  From our picture of classical
particle production it follows that the multiplicity around $y_1$
should be the same as around $y_2$ up to statistical fluctuations.

If we applied this analysis to an average event, the result above
would be trivial.  On the other hand, if we look at an event where the
fluctuation is very rare in the bin around $y_1$, we predict that
there will also be the same rare fluctuations around $y_2$!  Rare
events tend to fluctuate together over a wide range of rapidity!

A measurement of the width of the correlation length tells us
something about the underlying classical dynamics and is interesting
in itself.

First we note that in a nuclear collision $\frac{d N}{d y}$ is a
function of rapidity $y$, impact parameter of the nuclei $b$, and of
the configuration of color charges in the nuclei in this particular
collision, which we will symbolically denote $\rho$:
\be
\frac{d N}{d y} = \frac{d N}{d y} (y, b, \rho).
\ee
If one measures $\frac{d N}{d y}$ in a number of events and then takes
the average value at a fixed impact parameter of the colliding nuclei
$b$ the result should correspond to the averaging of the theoretical
prediction for $\frac{d N}{d y}$ over all configurations of color
charges in the colliding nuclei $\rho$, which we write as
\be
\left< \frac{d N}{d y} (y, b, \rho) \right>_\rho.
\ee
Event by event fluctuations in $\frac{d N}{d y} (y, b,\rho)$ are
characterized by the variance of that quantity which we will define as
\be
V \left[ \frac{d N}{d y} \right] = \left< \left( \frac{d N}{d y}(y, b, 
\rho) \right)^2 \right>_\rho - \left< \frac{d N}{d y} (y, b, \rho) 
\right>_\rho^2.
\ee
Thus we can also introduce the correlation function of the numbers of
particles measured at two different rapidity points $y_1$ and $y_2$ in
several events with the same impact parameter (that is for several
configurations of color charges $\rho$). According to the standard
mathematical methods we define the correlation coefficient
\be \label{ccoef}
{\cal C} (b; y_1, y_2) = \frac{ \left< \frac{d N}{d y_1 d y_2}(b,
\rho) \right>_\rho - \left< \frac{d N}{d y_1} (b, \rho) \right>_\rho \, \left<
\frac{d N}{d y_2} (b, \rho) \right>_\rho }{ \sqrt{ V \left[ \frac{d N}{d y_1} 
\right] \, V \left[ \frac{d N}{d y_2} \right]}}.
\ee
We can neglect the $y$ dependence of $\frac{d N}{d y} (y, b, \rho)$
for the rapidity intervals of the size $1/\alpha$. That way one can
see that when $| y_1 - y_2 | \lsim 1/\alpha$ we can neglect the
variation of $\frac{d N}{d y} (b, \rho)$ with $y$, which is purely
statistical in that interval, and put $\frac{d N}{d y_1} \approx
\frac{d N}{d y_2}$, which leads to ${\cal C} (b; y_1, y_2) = 1$. This is
the prediction of the classical emission picture discussed above and
in \cite{mv,yv}: the correlation function ${\cal C} (b; y_1, y_2)$
should be close to one for $| y_1 - y_2 | \lsim 1/\alpha$. When $| y_1
- y_2 | \gsim 1/\alpha$ the $y$ dependence in $\frac{d N}{d y}$ in the
interval between $y_1$ and $y_2$ becomes important and $\frac{d N}{d
y_1}$ and $\frac{d N}{d y_2}$ become uncorrelated, which would lead to
${\cal C} (b; y_1, y_2)$ being smaller than one.

One might note that in the leading powers of $A$ approximation both
numerator and denominator in the expression for ${\cal C} (b; y_1,
y_2)$ given by Eq. (\ref{ccoef}) are zero. The non-zero terms appear
once we include the corrections which are subleading in $A$. The first
non-zero terms in the numerator and denominator of Eq. (\ref{ccoef})
are suppressed by $A^{2/3}$ compared to the leading contribution in,
for instance, $\frac{d N}{d y_1 d y_2}(b, \rho)$. That is if we
calculate $\frac{d N}{d y_1 d y_2}(b, \rho)$ in the classical
approximation we would obtain the leading term in $A$ given by
Eq. (\ref{cor}), which is canceled in the numerator of
Eq. (\ref{ccoef}), and the subleading terms, which are suppressed at
least by $A^{2/3}$, but still arise from the classical
approximation. The largest of these subleading terms is suppressed by
exactly $A^{2/3}$ and results from the situation when both particles
at $y_1$ and $y_2$ are emitted from the same nucleon. Even though it
is subleading in $A$ this term is nevertheless larger than the
correlations induce by quantum corrections connecting the fields that
produce the particles at $y_1$ and $y_2$, since the latter are also
suppressed by extra powers of $\alpha_s$. The quantum correction
within one of the fields do not violate the factorization of
Eq. (\ref{cor}) and, therefore, cancel in the numerator of
Eq. (\ref{ccoef}). Thus the fact that the correlator in
Eq. (\ref{ccoef}) is equal to $1$ because of the terms in the
numerator and denominator which are subleading in $A$ does not
influence the fact that the correlations are classical.

In order to understand Eq. (\ref{ccoef}) better let us consider
different possible definitions of the correlation function. For
rapidity correlation function defined as
\be\label{r1}
{\cal R}_{\rho} (y_1, y_2) = \,\, \frac{ \frac{d N}{d y_1 d y_2 }(b,
\rho)\ - \frac{d N}{d y_1} (b, \rho) \, \frac{d N}{d y_2} (b,
\rho) }{ \frac{d N}{d y_1} (b, \rho ) \, \, \frac{d N}{d y_2} (b,
\rho)}
\ee
we expect that
\be
 {\cal R}_{\rho} (y_1, y_2) = O\left( A^{- \frac{2}{3}}\right).
\ee
Note, that in Eq.(\ref{ccoef}) we fix the value of the impact
parameter (transverse distance between centers of colliding nuclei)
which is also a characteristics of the event. The $A^{-
\frac{2}{3}}$ - corrections stem from region of integration $p_T
\,\approx\,q_T $ in Eq. (\ref{aaaa}). This is the subleading correction 
to the factorization of \eq{cor}. It could also be viewed as resulting
from the contribution when the two fields of Fig. \ref{eve} share the
same source of color charge. That term would definitely be suppressed
by $A^{2/3}$.  Thus the correlation coefficient of Eq. (\ref{r1})
defined without averaging over events does not reflect the
correlations that we are interested in. However, $R_\rho$ is not the
correlation coefficient usually employed in the experiments.  One
defines the correlation function ${\cal R}_{average}$ by
\be \label{r2}
{\cal R}_{average} (y_1, y_2) = \,\, \frac{\frac{1}{\sigma_{tot}} \left<
\frac{d \sigma}{d y_1\,\,d y_2}\ \right>_{b, \rho} - \frac{1}{\sigma_{tot}} 
\left< \frac{d
\sigma}{d
y_1} \right>_{b, \rho} \,\frac{1}{\sigma_{tot}} \left< \frac{d
\sigma}{d y_2}
\right>_{b, \rho} }{ \frac{1}{\sigma_{tot}}\left< \frac{d \sigma}{d y_1} 
\right>_{b, \rho} \, \, \frac{1}{\sigma_{tot}}\left<
 \frac{d \sigma}{d y_2} \right>_{b, \rho}}
\ee
where $\sigma_{tot}$, $\frac{d \sigma}{d y}$ and $\frac{d \sigma}{d
y_1\,d y_2}$ are total, single and double inclusive cross sections for
nuclear collisions that are defined by averaging over all events
($\rho$) and integrating over all impact parameters $(b_t)$. As was
shown long ago in the Pomeron approach function ${\cal R}_{average}$
generally is of the order of unity even in the case if have ${\cal
R}_{\rho} = 0$ \cite{LRN}. However, we would like to point out that
the main correlation which makes ${\cal R}_{average}$ large is the
simple correlation in impact parameters which has a very simple
underlying physics, which states that partons (gluons) are produced
independently but at the same value of impact parameter. In
McLerran-Venugopalan model one might expect that at high parton
densities or better to say in collisions of heavy nuclei the $b_t$ -
distribution for all physical observables are the same, namely,
$\Theta(R_A - b_t)$. In this case ${\cal R}_{average} \,\rightarrow
\,0$ showing a classical field emission even without the event
selection ($b_t$ fixing). If one does not integrate over the impact
parameter in Eq. (\ref{r2}) then without the impact parameter
correlations this correlation function will be small again failing to
reflect the type of correlations we are interested in.

The general picture of correlations for high parton density QCD turns out
to be very similar to pattern of correlations predicted by the Reggeon
approach which has been discussed in details for three decades
\cite{LRN,REGGEON,REG1,REG2,REG3,REG4,REG5,REG6,REG7,REG8,REG9,ms}. The 
first observation is that in both approaches Eqs. (\ref{aaaa}) and
(\ref{r1}) lead to conclusion that the secondary gluons (hadrons) are
originated from the independent production of clusters with the mean
multiplicity $ \left< N \right> = \int \frac{d N}{d y} (y, \rho, b_t
)\,\,d y $. Therefore, the probability to emit $k \times \left< N
\right>$ gluons or, in other words, $k$ clusters, is equal to
\be  \label{P1}
P_k\,\,=\,\,e^{- \left< N \right>}\,\,\frac{\left< N \right>^k}{
k!}\,\,.
\ee
Note, that \eq{P1} is a well known Poisson distribution.

The second observation stems from the classical field emission,
namely, $\frac{d N}{d y} (y, \rho, b_t ) = d = Const( y ) \gg 1 $ for
the rapidity interval of the order of $1/\alpha_s \gg 1 $.  We will
discuss why $d \gg 1$ below. Assuming this we can easily see that we
can neglect statistic fluctuations in $N( y,\rho, b_t)$, and,
therefore, $\left< N \right> = Y \, d $ where $Y$ is the accessible
rapidity interval ($Y \propto 1/\alpha_s$ ). We can claim even that
$N(y,\rho,b_t)$ which is defined as the number of gluons ( hadrons) in
the rapidity interval $ y - \Delta y \div y + \Delta y$ is equal to
\be \label{P2}
N(y,\rho,b_t)\,\,=\,\,k \times d\, \times\,2 \Delta y\,\,.
\ee
Therefore, measuring $N(y,\rho,b_t)$ we fix the number of clusters $k$
or, in other words, we select a configuration which is produced with
probability $P_k$ (see \eq{P1}).

Let us recall the main predictions that will hold both in classical
field approach ( McLerran-Venugopalan model ) and in the Reggeon
description. If we select the events where the multiplicity of the
particles in the rapidity bin $y_1 - \Delta y \div y_1 + \Delta y $
around $y_1$ ($N (y_1)$) is fixed and average the multiplicity in the
rapidity interval $y_2 - \Delta y \div y_2 +
\Delta y $ over these events we predict that it will be equal to
\be
\left< N ( y_2 ) \right>_{N(y_1) \, fixed} \,\,\,=\,\,N(y_1)\,\,.
\ee
Also the number of neutral pions in rapidity interval $y_1 - \Delta y
\div y_1 + \Delta y $ ($N_0 (y_1)$) averaged over the events with a
fixed number of charged pions in the same bin should be equal to
\be 
\left< N_0 (y_1) \right>_{N^{ch}(y_1) \, fixed} \,\,\,=\,\,\, 
\frac{1}{2} \,\,\,N^{ch} (y_1)\,\,,
\ee
where $N^{ch} (y_1)$ is the number of charged pions. Both models
predict that for heavy ion-ion collisions in the kinematic region of
saturation\cite{mv,glr} we have
\be
{\cal R}_{average} (y_1, y_2)\,\,\,\longrightarrow\,\,\,0\left(
\frac{1}{A^{2/3}}\right)\,\,.
\ee
One can see that all these prediction are direct consequences of
\eq{P1} and \eq{P2}. Much more detailed predictions could be found in Refs.
\cite{REGGEON,REG1,REG2,REG3,REG4,REG5,REG6,REG7,REG8,REG9,ms}.

The high density QCD and soft Reggeon approaches can be distinguished
by measuring the transverse momenta distributions or/and transverse
momenta correlation between produced particles. Indeed, in high parton
density QCD the typical transverse momentum of produced gluons
(saturation momentum $Q_s (A, x)$ ) is large and depends on $A$
\cite{mv,glr} while in the Reggeon approach this momentum is constant
and rather small --- about 2 GeV, which results from the slope
estimates of the soft pomeron. However, at first sight, this main
difference does not influence the shape of the rapidity correlations
and only increases the multiplicity of produced particles making our
predictions more reliable in this case. Indeed, the saturation of the
gluon density leads to $d \propto Q^2_s(W, A)$
\cite{mv,yv,JKMW,glr,me,LT} where $W$ is the energy in the center of mass 
frame and $Q^2_s(W, A)$ is the saturation scale \eq{sat}, which is
similar to $\Lambda$ of \eq{lam}.  Since $Q^2_s(W,A)$ increases with
$W$ and $A$ (at least $Q_s^2 \propto
A^{\frac{1}{3}}$\cite{mv,yv,JKMW,glr,me} but it could even be
proportional to $A^{\frac{2}{3}}$ \cite{LT}) we expect $d \gg 1 $
which makes all of our estimates much more accurate than the results
of the Reggeon approach \cite{REG5} \cite{TA}. As far as the shape of
the rapidity correlations is concerned the saturation of the parton
densities \cite{mv,yv,JKMW,glr,me,LT} leads to sufficiently large
correlation length which is proportional to $1/\alpha_s (Q_s)$. For
heavy nuclei and/or high energies $Q_s$ increases and $\alpha_s
\rightarrow 0$. This fact leads to \eq{P2} being better justified in
the high parton density QCD than in the case of the Reggeon approach.

\section{Conclusions}

In this paper we have considered the consequences of the model of
classical emissions on the particle production mechanisms in heavy ion
collisions \cite{mv}. We have argued that the classical fields do not
vary significantly over the rapidity intervals of the order of
$1/\as$. This led us to conclude that the multiplicities of the
particles produced in the early stages of the collisions (gluons) are
correlated over $1/\as$ units of rapidity on the event-by-event
basis. We then gave an argument demonstrating that the number of
gluons generated in a particular collision is proportional to the
number of pions produced in the final state. Therefore the correlation
in the multiplicities of the produced gluons would reflect themselves
in the multiplicities of the final state pions. We have then
constructed the correlation function ${\cal C} (b; y_1, y_2)$ in
\eq{ccoef}, which could be measured experimentally and is equal to $1$ 
when the multiplicities of particles at the rapidities $y_1$ and $y_2$
are correlated and is less than $1$ otherwise. Surprisingly this
effect comes from the terms in the numerator and denominator of the
expression for ${\cal C} (b; y_1, y_2)$ given by \eq{ccoef} which are
subleading in $A$. We predict ${\cal C} (b; y_1, y_2)$ to be close to
$1$ when $|y_1 - y_2| \le 1/\as$ and fall off for larger rapidity
intervals.

We have also used the model of classical particle production to
explain KNO scaling. In a simple toy model which resulted from trying
to mimic the main features of the classical emissions we have
reproduced KNO scaling for the multiplicities of produced
particles. The result is given in \eq{knof}. Even though the exact
shape of the KNO distribution function is, probably, different from
our toy model prediction, we believe that the model captures its main
features.  For the case of nuclei we predict the KNO function to
depend on the atomic number $A$ in addition to the usual dependence on
$N/{\overline N} (y)$. We predict that at very high energies KNO
scaling in nuclear collisions will be violated due to the running of
the coupling constant, which would get smaller as parton density
increases.

The main results of this paper can be summarized in the following way:

\begin{enumerate}
\item\,\,\, We have derived KNO scaling from the classical emission 
picture of particle production (see Eqs. (\ref{knoff}) and
(\ref{knof})).

\item\,\,\, We predict rapidity correlations on the scales 
$\delta y \sim 1/\as$ in the particle (pion) production in heavy ion
collisions on the event-by-event basis (Sect. IV).

\item\,\,\, We proposed a correlation coefficient ${\cal C} (b; y_1, y_2)$ 
which would allow one to measure the predicted correlations (see
\eq{ccoef}).

\end{enumerate}

\section*{Acknowledgments}

The authors would like to acknowledge helpful and encouraging
discussions with Errol Gotsman, Miklos Gyulassy, Jamal
Jalilian-Marian, Uri Maor, Al Mueller, Mark Strikman, Kirill Tuchin,
Raju Venugopalan, and Heribert Weigert.

This work was carried out while E.L. was on sabbatical leave at
BNL. E.L. wants to thank the nuclear theory group at BNL for their
hospitality and support during that time.

The research of E.L. was supported in part by the Israel Science
Foundation, founded by the Israeli Academy of Science and Humanities.

This research has been supported in part by the joint
American--Israeli BSF Grant $\#$ 9800276. This manuscript has been
authorized under Contract No. DE-AC02-98CH10886 with the
U.S. Department of Energy.

\end{document}